\documentclass[12pt]{article}

\newread\testifexists
\def\GetIfExists #1 {\immediate\openin\testifexists=#1
    \ifeof\testifexists\immediate\closein\testifexists\else
    \immediate\closein\testifexists\input #1\fi}

\usepackage{gthstyle}
\immediate\openin\testifexists=e
    \ifeof\testifexists\immediate\closein\testifexists\else
    \immediate\closein\testifexists\input e\fipsf

\def\Bbb#1{\setbox0=\hbox{$\tt #1$}  \copy0\kern-\wd0\kern .1em\copy0}

\def\bbf#1{\setbox0=\hbox{$#1$} \kern-.025em\copy0\kern-\wd0
        \kern.05em\copy0\kern-\wd0 \kern-.025em\raise.0433em\box0}

\immediate\openin\testifexists=a
    \ifeof\testifexists\immediate\closein\testifexists\else
    \immediate\closein\testifexists\input a\fimssym.def          

\def\a{\alpha}      \def\b{\beta}   \def\g{\gamma}      
\def\d{\delta}      \def\D{\Delta}  \def\e{\varepsilon}
               \def\L{\Lambda}
\def\m{\mu}         \def\f{\phi}            
\def\n{\nu}         \def\j{\psi}    
\def\r{\varrho}       \def\SS{\Sigma}

 \def\LL{{\cal L}} 
\def\pa{\partial} \def\ra{\rightarrow}

\def\dd{{\rm d}}

\def\fract#1#2{{\textstyle{#1\over#2}}}
\def\ffract#1#2{\raise .3 em\hbox{$\scriptstyle#1$}\kern-.25em/
                \kern-.2em\lower .2 em \hbox{$\scriptstyle#2$}}

\def\half{\fract12} \def\quart{\fract14} 

\def\part#1#2{{\partial#1\over\partial#2}}

\newcommand{\be}{\begin{eqnarray}}
\renewcommand{\le}[1]{\label{#1}\end{eqnarray}}
\newcommand{\ee}{\end{eqnarray}}
\newcommand{\eqn}[1]{(\ref{#1})}

\newcommand{\fn}{\footnote}
\newcommand{\newsec}[1]{\section{#1}\setcounter{equation}{0}}

\begin{document}

\begin{titlepage}
\begin{center}
\hfill ITF-2002/39  \\ \hfill SPIN-2002/22  \\ \hfill {\tt
hep-th/0207179}\\ \vskip 20mm

{\Large{\bf PERTURBATIVE CONFINEMENT}\fn{Presented at QCD'02,
Montpellier, 2-9th July 2002.}}

\vskip 10mm

{\large\bf Gerard 't~Hooft}

\vskip 4mm Institute for Theoretical Physics \\  Utrecht
University, Leuvenlaan 4\\ 3584 CC Utrecht, the Netherlands
\medskip \\ and
\medskip \\ Spinoza Institute \\ Postbox 80.195 \\ 3508 TD
Utrecht, the Netherlands
\smallskip \\ e-mail: \tt g.thooft@phys.uu.nl \\ internet:
\tt http://www.phys.uu.nl/\~{}thooft/

\vskip 6mm

\end{center}

\vskip .2in
\begin{quotation} \noindent {\large\bf Astract } \medskip \\
A Procedure is outlined that may be used as a starting point for
a perturbative treatment of theories with permanent confinement.
By using a counter term in the Lagrangian that renormalizes the
infrared divergence in the Coulomb potential, it is achieved that
the perturbation expansion at a finite value of the strong
coupling constant may yield reasonably accurate properties of
hadrons, and an expression for the string constant as a function
of the QCD \(\L\) parameter.
\end{quotation}

\vfill \flushleft{\today}

\end{titlepage}

\newsec{Introduction}
In recent work\cite{GrTh}, Greensite and Thorn described a simple
procedure to produce wave functions for states that may describe
stringlike features for hadrons in QCD. As an {\it Ansatz}, a
wave function was taken of the form \be \j(\vec x_1,\,\vec
x_2,\cdots, \vec x_{N-1})=A\prod_{i=1}^N \f(\vec u_i)\ ,\le{wave}
where \be \vec u_i=\vec x_i-\vec x_{i-1}\quad,\qquad \vec x_0,\
\vec x_N\quad \hbox{fixed.}\le{wave1} They then propose to use a
variational principle: \(\f\) is chosen such that the total energy
\(\cal E\), defined by \be {\cal E}=\sum T^{\rm kin}+\sum V^{\rm
Coulomb}\ ,\le{energy} is minimized. Here, \(V^{\rm Coulomb}\) is
taken to be simply the Coulomb potential, \be V^{\rm
Coulomb}=\sum_i V^{\rm C}(\vec u_i)\quad ,\qquad V^{\rm C}(\vec u)
= -{\a_s\over |\vec u|}\ .\le{Coulomb} Subsequently, it is
proposed to try ``improved {\it Ans\"atze}", and it is suspected
that those wave functions that give a string like appearance will
carry the lowest amounts of total energy \(\cal E\).

\begin{figure}[b] \setcounter{figure}{0}
\begin{center} \epsfxsize=160 mm\epsfbox{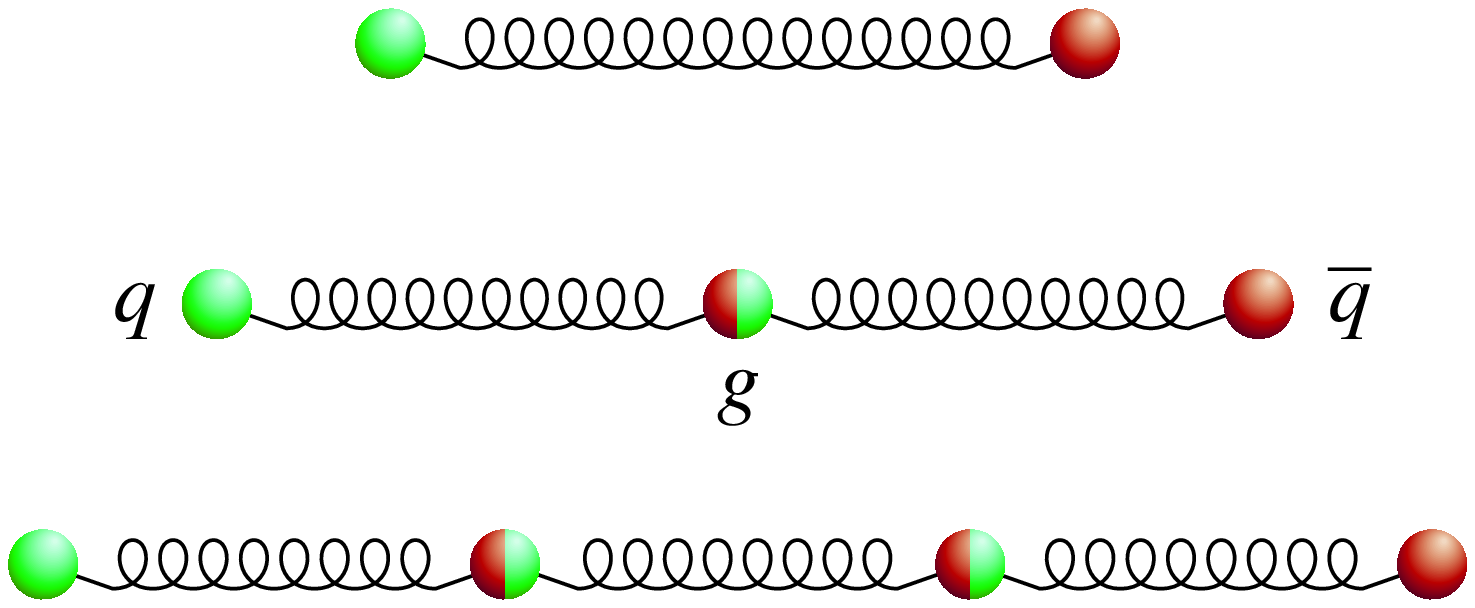}
  \caption\footnotesize{\footnotesize{The Gluon Chain Model}}
\end{center}
\end{figure}

Some aspects of this proposal, however, are less than
satisfactory. Certainly, permanent confinement is not at all a
guaranteed property. If we consider the complete set of elements
of Fock space, we must include all those states in which gluons
do not form a chain, and therefore it is to be expected that
\emph{unitarity} will not hold if we restrict ourselves to
chainlike states. A complete set of stringlike states could be
found by replacing the Coulomb potential by a confining
potential, but can one justify such a prescription in a theory
where the classical limit would not exhibit any resemblance to a
confinement mechanism?

In this paper, it is attempted to answer this question.
Confinement should be regarded as a natural renormalization
phenomenon in the infrared region of a theory. The way to deal
with renormalization counter terms in the infrared is not so
different from the more familiar ultraviolet renormalization. In
Section~\ref{sec:renorm}, we explain how to look upon bare
Lagrangians, renormalized Lagrangians, and how counter terms may
switch position in the perturbation expansion.

In Section~\ref{sec:Coulomb}, the counter term expected for
renormalizing the Coulomb potential is discussed. At this stage
of the procedure, fixing the gauge condition to be the radiation
gauge is crucial. There are no ghosts then, so unitarity of the
procedure is evident, as long as we take for granted that the
\emph{ultraviolet} divergences can be handled separately by the
usual methods (the ultraviolet nature of the theory is assumed
not to be affected by what happens in the infrared). We then
explain how to work with such counter terms
(Section~\ref{sec:IR}).

Working with non-gauge-invariant, non-Lorentz-invariant
expressions will be cumbersome in practice, which is why we seek
for a more symmetric approach. In Section~\ref{sec:gaugeinv}, it
is shown that gauge-invariant renormalization effects in the
infrared can also easily lead to confinement. Again, the
renormalization counter terms are essentially local but soft. they
cannot exist as \emph{fundamental} interaction terms in a
quantized field theory, but they \emph{must} arise as effective
interactions due to higher loop contributions.

Technicalities are worked out in Section~\ref{sec:Legendre}, and
finally, it is shown that magnetic confinement could be handled
in a similar fashion, although, of course, the standard Higgs
mechanism is a more practical way to deal with such theories.

\newsec{Renormalization\label{sec:renorm}}
To illustrate our new approach, we remind the reader of the
standard procedures for ultraviolet renormalization. In that case,
the ``bare Lagrangian", \(\LL^{\rm bare}\), is rewritten as
\be\LL(g,\,A,\cdots)=\LL^{\rm bare}+\D\LL-\D\LL\ ,\le{Lagr} where
one subsequently interpretes the combination \be\LL^{\rm
renorm}(g^{\rm ren},\,A^{\rm ren})=\LL^{\rm bare}+\D\LL \
,\le{Lren} as the finite, ``renormalized" Lagrangian. Using a
regulator procedure such as, for instance, a lattice, one may
allow that \(\LL^{\rm bare}\), the bare couplings \(g\) and the
bare fields \(A\) become ill-defined, or "infinite", in the limit
where the regulator vanishes (\(a\downarrow 0\)), but the limit
should be arranged in such a way that the renormalized coupling
parameters \(g^{\rm ren}\) are finite, and the renormalized fields
\(A^{\rm ren}\) are finite operators.

\(\D\LL\) does contain infinities, but these are precisely
constructed so as to cancel all infinities that one encounters in
calculating the higher loop corrections, order by order in
perturbation theory. In calculating these higher order
corrections, one again uses \(\LL^{\rm renorm}\), not \(\LL^{\rm
bare}\), because, indeed, \(\D\LL\) is again needed to cancel the
subdivergences at higher orders, as one can verify explicitly.

It is important to note now that indeed we could have chosen
anything for \(\D\LL\), as long as we insist that whatever we
borrow to include in the ``lowest order" Lagrangian \(\LL^{\rm
renorm}\), we return when a higher loop correction is computed.
The procedure pays off in particular if we manage this way to
keep the higher order corrections, after having returned
\(\D\LL\) together with them, as small as possible.

What does ``as small as possible" mean? This is something one may
dispute. The \emph{renormalization group} is based on the
observation that when features are computed for external momenta
(and possible relevant mass terms) at a typical scale \(\m\), then
the optimal choice for \(\D\LL\) depends on this scale \(\m\). For
small scale transformations, the transition from one choice of
\(\D\LL\) to another may not seem to be very important, but for
transitions from one scale to a very much larger or smaller one,
it is. The result is that the renormalized coupling strength
\(\a_s\) depends on \(\m\), and it may vanish (logarithmically) as
\(\m\ra\infty\).

There appears to be no objection against a wider use of such a
procedure. If our higher order amplitudes exhibit infrared
divergences, besides the ultra-violet ones, we could absorb them
in \(\D\LL\) as well. This time, \(\D\LL\) is not expected to
affect the coupling strengths and the field operators, but a
quantity that is ideally suited to be renormalized by such terms
is the effective Coulomb potential. Thus, after already having
dealt with the ultraviolet divergences, we add to the ``lowest
order Lagrangian" \(\LL^{\rm renorm}\), a further term \(\D\LL\)
that affects the Coulomb potential. We just borrow this term from
the higher order corrections. As soon as these are calculated, we
will be obliged to return the loan.

\newsec{The Renormalized Coulomb potential\label{sec:Coulomb}}

In the sequel, we refer to the usual QCD Lagrangian as \(\LL^{\rm
bare}\), even if the ultraviolet divergences have already been
taken care of. \(\D\LL\) will then only describe the new,
infrared counter terms. Thus, we write \be \LL^{\rm bare}=-\quart
F_{\m\n}F_{\m\n}-\overline{\j}(\g D+m)\j+\LL^{\rm
gauge-fix}+\LL^{\rm ghost}+\D\LL-\D\LL\,.\le{Lbare} The
\emph{radiation gauge}, \be\pa_i A_i=0\ ,\le{radgauge} is useful
if questions of unitarity are to be kept under control, such as
here. As usual, Latin indices \(i,\,j,\cdots\) are taken to run
from 1 to 3, whereas Greek indices \(\m,\,\n,\cdots\) must run
from 0 to 3 (or from 1 to 4). The color indices are of no direct
concern at this stage, since all our expressions at this order
are diagonal in the color indices, which is why we suppress them,
for the time being.  In this radiation gauge, the kinetic part of
the gauge field Lagrangian will be \be \LL^{\rm
bare}=-\half(\pa_\m
A_\n)^2+\half(\pa_\n A_\n)^2 =
-\half(\pa_i A_j)^2+\half(\pa_0A_j)^2+\half(\pa_jA_0)^2
\,.\le{Lbare1} Here, \(A_i\) behaves as an ordinary dynamical
variable, except for the constraint~\eqn{radgauge}, whereas
\(A_0\) only emerges in a spacelike derivative term; its timelike
derivative, \(\pa_0A_0\) is absent (it cancelled out in
\eqn{Lbare1}). Therefore, \(A_0\) generates the instantaneous
Coulomb interaction between all color charges.

The Coulomb interaction resulting from the bare Lagrangian is \be
V^{\rm Coulomb}_{q,\overline{q}}(\vec r\,)=-{\a_s\over|\,r|}\
.\le{VCoulomb} Our procedure will be particularly important when
infinite infrared corrections show up. This is exactly what we
expect in a theory with confinement. We do not intend to ``prove"
permanent confinement here. Such a `proof', or at least some
justification for its assumption, is assumed to be given
elsewhere.\cite{GtHconf} We take for granted now that the
renormalized Coulomb potential will be a confining one.
Typically, what is needed is an `effective' potential that is
diagonal in the color indices, and takes up the new form \be
V^{\rm ren}_{q,\overline q}(\vec r\,)=-{\a_s\over|\,r|}+\r|\,r|\
,\le{Vren} where \(\r\) is an effective string constant. Notive
that the signs here are chosen in such a way that the Coulomb
\emph{force} keeps a constant sign. The presence of a zero in
Eq.\eqn{Vren} is of no significance.

The familiar Coulomb term, Eq.\eqn{VCoulomb}, obeys the Laplace
equation \be\vec\pa^{\,2} V^{\rm Coulomb}(\vec
r)=4\pi\a_s\,\d^3(\vec r)\ ,\le{LaplC} whereas the extra term in
Eq.\eqn{Vren} obeys the Laplace equation \be\vec\pa^{\,2} (\r
|\,r|)={2\r\over|\,r|}\quad;\qquad
\vec\pa^{\,2}\left({2\r\over|\,r|}\right)=-8\pi\r\,\d^3(\vec r)\ .
\le{Laplstr} Therefore, in 3 dimensional Fourier space, the
renormalized Coulomb potential is\cite{KogutSusskind} \be V^{\rm
ren}(\vec k)=-{4\pi\a_s\over \vec k^2}-{8\pi\r\over\vec
k^4}\,,\le{Fourier} so that the Laplace equation for the
renormalized potential will have to be constructed from\be{\vec
k^4\over\vec k^2+2\r/\a_s}\,V^{\rm ren}=-4\pi\a_s\d^3(\vec r
)\le{Laplren} (Here, an admittedly ugly mixed Fourier-coordinate
space notation was admitted). Because of the extra term
\(2\r/\a_s\), the renormalized Lagrangian should contain the
following, modified Coulomb term:\be\LL^{\rm
bare}+\D\LL=\dots+\half\pa_jA_0\left(1-{2\r/\a_s\over\vec
k^2+2\r/\a_s}\right)\pa_jA_0\ ,\le{LCoulren} so that \(\D\LL\)
must be chosen to be \be\D\LL=-\half\pa_jA_0\,{2\r/\a_s\over\vec
k^2+2\r/\a_s}\,\pa_jA_0\ .\le{DeltaL}

In coordinate space, this takes the form \be\int\D\LL(\vec x,t)
\,\dd^3\vec x =-\half\int\dd^3\vec x\int\dd^3\vec
x'\,\pa_jA_0(\vec x) \, G(\vec x-\vec x')\,\pa_jA_0(\vec x')\
,\nonumber\\ {\rm where}\qquad G(\vec
r)={8\pi\r\over\a_s}\,{1\over |\,r|}\,
e^{-\sqrt{2\r/\a_s}\,|\,r|}\ . \le{DS} Note that \(G(\vec x-\vec
x')\) is fairly `soft', both in  the infrared (where it decays
exponentially), and in the ultraviolet (where it is integrable and
considerably weaker than the Dirac delta function of the canonical
term). On the other hand, this term cancels the canonical term
\emph{completely} at \(\vec k\ra 0\).

\begin{figure}[t]
\begin{center} \epsfxsize=130 mm\epsfbox{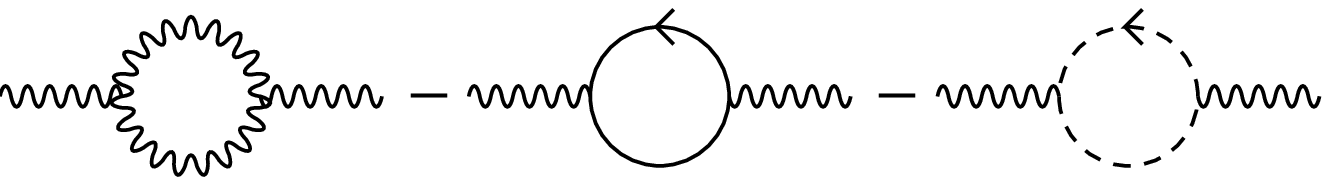}
  \caption{\footnotesize{The one-loop corrections to the gluon propagator: the gluon,
  the quark and the ghost term (the minus signs are due to the
  Fermi statistics of the last two; the
  last diagram, the ghost, actually does not
  contribute in the radiation gauge).}}
  \label{fig:canloop}
\end{center}
\end{figure}

\newsec{Infinite infrared renormalization\label{sec:IR}}
In the standard perturbative approach, the one-loop correction to
the gluon self-energy is infrared divergent. The self-energy
correction from the diagrams of the type of
Figure~\ref{fig:canloop} is \be C\,g^2(k^2\d_{\m\n}-k_\m
k_\n)\log{1\over k^2}\ .\le{canampl} It would produce a correction
term to the Coulomb potential of the form \(\log(1/\vec k^2)\)
times that term. The divergence at \(|\vec k|\ra\infty\) is the
standard type that is to be taken care of by ultraviolet
renormalization. It is the divergence near \(|\vec k|\ra 0\),
which goes as \(2C\log(1/|\vec k|\), that we now wish to regulate.
The counter term produced by Eqs.~\eqn{DeltaL} and \eqn{DS} has
the form \be -{\m^2\over \vec k^{\,2}+\m^2}(k^2\d_{\m\n}-k_\m
k_\n)\quad;\qquad \m^2\equiv 2\r/\a_s\ .\le{counterterm} The
coefficient \(\m\) could now be required to be adjusted in such a
way that Eq.~\eqn{counterterm} cancels Eq.~\eqn{canampl} ``as well
as possible".

\begin{figure}[t]
\begin{center} \epsfxsize=130 mm\epsfbox{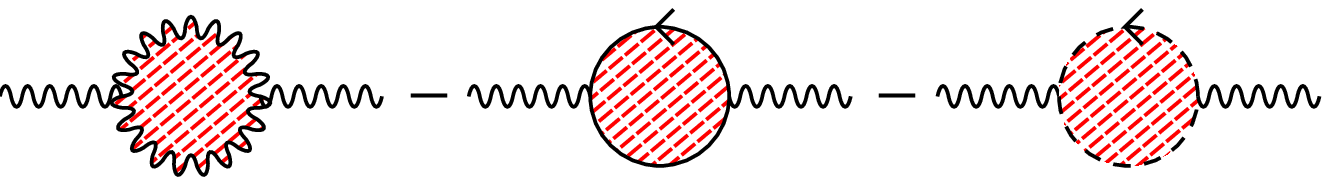}
  \caption{\footnotesize{The one-loop corrections to the gluon propagator
  in the presence of a confining potential.}}
  \label{fig:confloop}
\end{center}
\end{figure}

\begin{figure}[h]
\begin{center} \epsfxsize=130 mm\epsfbox{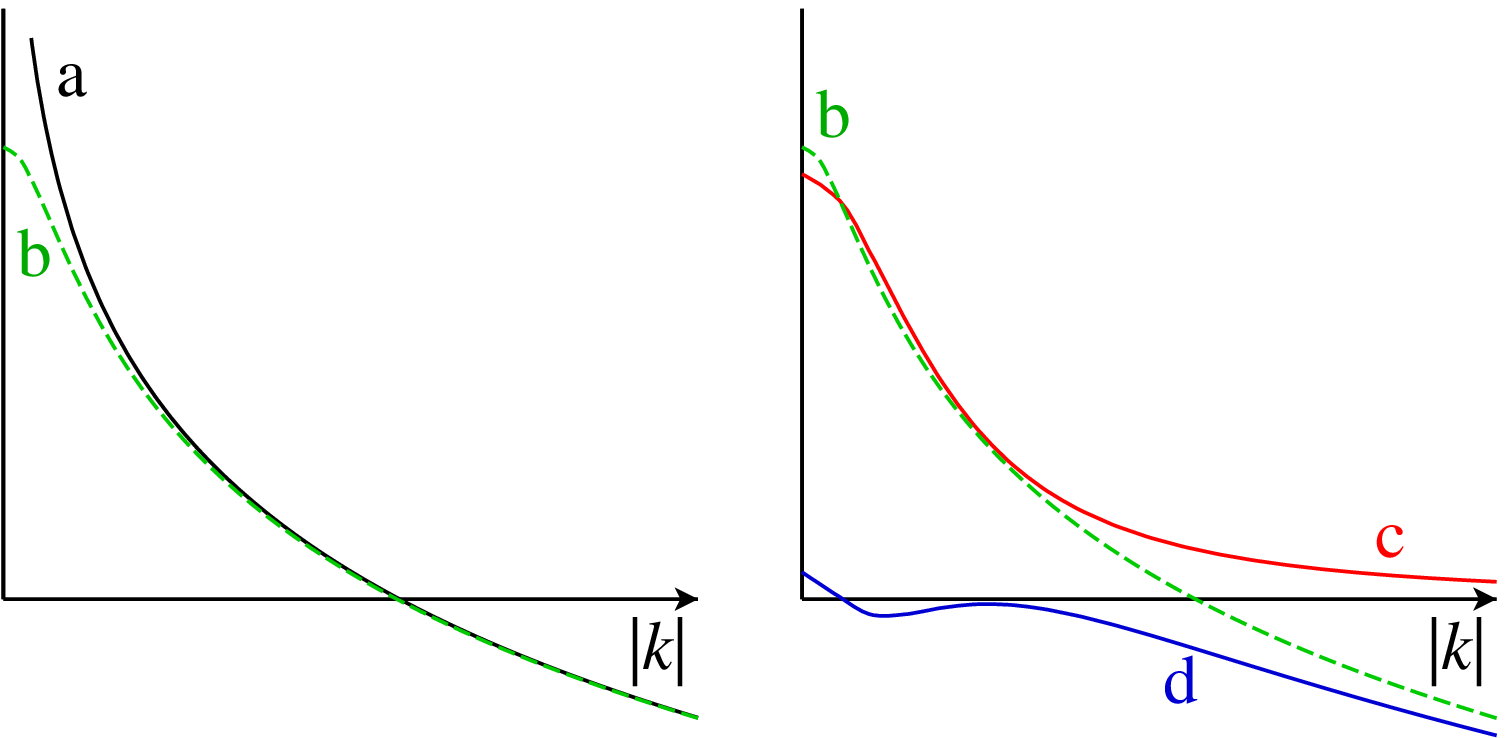}
  \caption{\footnotesize{The one-loop corrections to the gluon potential. $a$) The standard
  pure logarithm (in the case of massless flavors only). $b$) Estimate of the
  modification due to the confinement of the gluons inside the loop; the
  singularity at \(k\ra 0\) is exected to disappear. $c$) The counter term,
  Eq.~\eqn{counterterm}. $d$) The expected total one-loop correction, after subtraction of the
  counter term.}}
  \label{fig:SEcurves}
\end{center}
\end{figure}

Actually, the first order correction is not quite
Eq.~\eqn{canampl}, because the first order Lagrangian was
modified by the \(\D\LL\) term. The diagram that has to be
calculated actually is the one of Figure~\ref{fig:confloop},
where the shaded area denotes the effects from the confining
potential, Eq.~\eqn{Vren}. The effects this is expected to have on
calculating the higher order corrections to the Coulomb potential
are sketched in Figure \ref{fig:SEcurves}.

Although the counter term \eqn{DeltaL} is sufficient to produce
the above effects at higher order, one could also think of
replacing it by some gauge-invariant, or nearly gauge-invariant
expression, such as \be\D\LL=+\quart\int F_{\m\n}(\vec x,t)G(\vec
x-\vec x')F_{\m\n}(\vec x',t)\,\dd^3\vec x'&\ ;\label{DeltaL1}\\
G(\vec r)={8\pi\over \a_s}\,{1\over|\vec
r\,|}\,e^{-\sqrt{2\r/\a_s}{|\vec r\,|}}&\ .\le{Green} Notice that
this is only gauge-invariant in the Abelian case. In non-Abelian
theories, \(G\) must be replaced by a gauge-invariant path-ordered
expression.

On the other hand, in Eq.~\eqn{Green}, one could actually allow
for more general expressions such that the higher order
corrections to the Coulomb potential cancel out completely. This
would \emph{not} turn \emph{all} higher order corrections to zero,
since we limit ourselves to counter terms that are strictly local
in time (see Eq.~\eqn{DeltaL1}.

Note that, in all cases, the \(\D\LL\) that was inserted at zeroth
order, is borrowed from the correction at higher order (typically
the one-loop order), where it is returned, so as to render these
correction terms as small as possible. There is considerable
freedom in choosing \(\D\LL\), allowing it to be non-local in
space, but we do insist that it stays local in time. This will
guarantee that the Lagrangian we work with will always produce
unitary $S$-matrices at all stages. It also guarantees that, if
the \emph{first} loop corrections are small, also the higher loop
corrections stay small, just like the situation we have in
conventional renormalization procedures, where this can be proven
using dispersion relations.\cite{DIAGRAMMAR}

\newsec{Gauge-invariant counter terms\label{sec:gaugeinv}}
The procedure described in the previous section had the
disadvantage that gauge-invariance is lost. Locality of the
counter term in time, and the fact that it is cancelled at higher
order, should minimize the damage done, but still it might be
preferable to keep the infrared counter terms gauge-invariant
from the start. Eq.\eqn{DeltaL1} is a step in the right
direction, but then, the non-local kernel~\eqn{Green} should also
be made gauge-invariant, which is somewhat complicated in
practice. It would be easier if we could fabricate the counter
terms in such a way that the lowest order Lagrangian obtains the
form \be \LL^{\rm renorm.}\ =\ \LL^{\rm bare}+\D\LL\ =\ f(-\quart
F_{\m\n}F_{\m\n})\,.\le{function}

We have not yet worked out the formal scheme of correcting for
this modification at higher orders, but it might be doable. The
point that will be made in this section is that, with certain
minor modifications from the canonical theory \(f(x)=x\), which
only need to be sizable at small values of \(x\), theories can be
obtained that provide for rigorously confining potentials.

To demonstrate this, we first rewrite\cite{GtHstrong} the
Lagrangian~\eqn{function}, including a possible external source,
as \be \LL(A,\f)=-\quart Z(\f)F_{\m\n}F_{\m\n}-V(\f)+J_\m
A_\m\,.\le{Lphi} As the field \(\f\) was given no kinetic term
here, the Lagrange equations simply demand that we extremize with
respect to the \(\f\) field, so that equivalence with
Lagrangian~\eqn{function} is easy to establish. In the following,
we consider the static situation, and by concentrating on one
color excitation only, we ignore the non-Abelian terms that
otherwise could arise.

To solve the static case, with fixed sources \(J_0\), the
complete equations are easier to handle if we first solve the
equations for the fields \(A_\m\). One gets
\be\pa_i D_i=\r=J_0 \ ;&\label{divD}\\
\vec D=Z(\f)\vec E\quad,\quad E_i=&-\pa_iA_0\ {}.\le{DE} By
adding to the Lagrangian \eqn{Lphi} a term \be -{1\over
2Z}(Z\pa_i A_0+D_i)^2\,,\le{extraL} where \(D_i\) is a free
variable, we see that what has to be extremized is the
Hamiltonian   \be{\cal H}\ =\ D_i\dot{A_i}-\LL\ =\ \half\,{\vec
D^2\over Z(\f)}+V(\f)+J_0A_0\ ,\le{Hamilton} which is just
\(-\LL\) in the static case,  with Eq.~\eqn{divD} as a constraint.

\begin{figure}[t]
\begin{center} \epsfxsize=110 mm\epsfbox{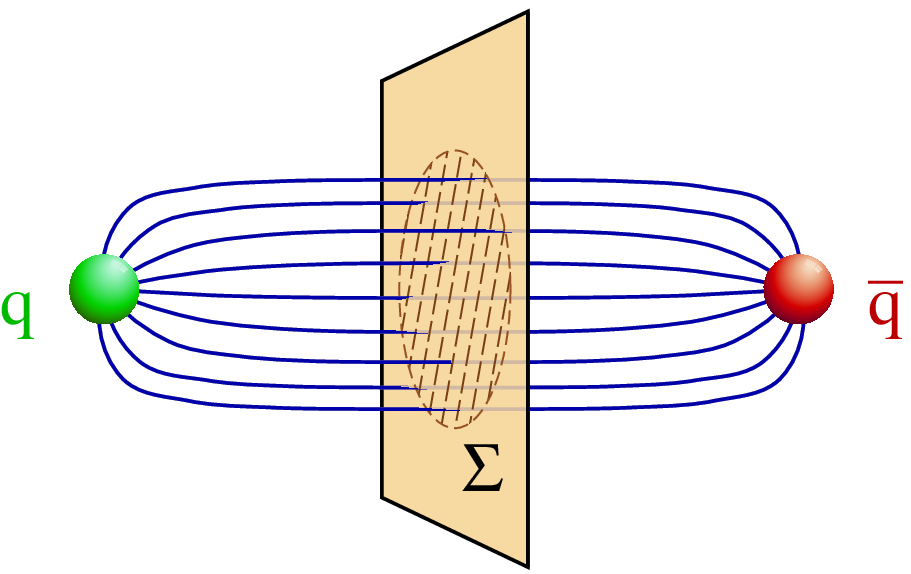}
  \caption{\footnotesize{The quark-antiquark vortes in a classical model. Shown
  are the \(\vec D\) field lines (whose divergence is controlled by the quark
  sources), and a sheet crosssecting these lines. Only inside the surface \(\SS\)
  (indicated as a shaded ellipse) they differ appreciably from zero.}}
  \label{fig:vortex}
\end{center}
\end{figure}

Let us assume momentarily that the minimum energy configuration
(in the presence of quark sources) is that of a vortex, as is
illustrated in Figure~\ref{fig:vortex}. The \(\vec D\) field
lines may spread over a surface area \(\SS\). In that case, the
predominant field strength \(\overline{D}\) is \be
\overline{D}\simeq {Q^{\rm
quark}\over\SS}\quad\ra\qquad\SS={Q\over\overline{D}}\
,\le{Dfield} where \(Q\) stands for the (Abelian) charge of a
quark. The energy per unit of length is \be{{\cal E }\over
L}\equiv\r^{\rm string}\quad=\quad\min_\SS\left(\SS\,U(\overline{
D})\right)\quad=\quad\min_D\left(Q{U(D)\over D}\right)\,.\le{rho}
If, indeed, Eq.~\eqn{rho} has a non-vanishing minimum, then
Fig.~{\ref{fig:vortex} correctly represents the configuration
with optimal energy. Note that, in the Maxwell case, one would
have: \be U(D)=\half\,D^2\ ,\qquad\hbox{(Maxwell)},\le{Maxwell}
so then there is no non-vanishing minimum, so that the field
lines spread out as usual. The minimum exists if, for low values
of \(D\), its energy goes as \be U(\vec D)\downarrow \r|\vec D|\
,\le{linear1} which, for simplicity, we take to be the case as
\(|\vec D|\ra 0\). For large \(\vec D\), we expect asymptotic
freedom to come into effect, so that the Maxwellian behaviour
\eqn{Maxwell} is resumed. For simplicity, also, we take \(Q=1\),
so that \(\r=\r^{\rm string}\).

Thus, we see that Eq.~\eqn{linear1} is sufficient to obtain the
effect of linear vortex formation. We now investigate what this
may imply for the function \(f(x)\) in Eq.\eqn{function}.

\newsec{An exercise in Legendre transformations\label{sec:Legendre}} First, one has
to establish the required relation between \(Z(\phi)\) and
\(V(\phi)\). extremizing with respect to \(\phi\) implies
\be\part{ }\phi \left({\half D^2\over
Z(\phi)}+V(\phi)\right)=0\quad\ra\quad\half D^2=- \part V{(1/Z)}\
  .\le{dphi} Regarding \(U\) as a function of \(\half D^2\) and
\(\phi\), we have \be \dd U=\half D^2\dd{1\over Z}+{1\over Z}\dd(
\half D^2)+\dd V={1\over Z}\,\dd(\half D^2)\,.\le{dU} In the
domain where we require Eq.\eqn{linear1} to hold, we therefore
have \be {1\over Z}={\dd U\over\dd(\half
D^2)}\quad\approx\quad{\dd(\r D)\over\dd \half D^2}={\r\over D}\
  .\le{ZD} From Eq.~\eqn{dphi}, one deduces \be V=-\int\half
D^2\,\dd\left({1\over
Z}\right)\quad\approx\quad\half\r\,D=\half\r^2 Z\ .\le{VZ} At
large values of \(D\), we expect the Maxwellian behaviour
\eqn{Maxwell}, so that there \be Z\ra 1\quad:\qquad \part
VZ\ra\infty\ .\le{largeD} Whether or not \(V\) itself tends to a
constant there, or to infinity, is not very important. All in all,
we expect the behaviour sketched in Fig.~\ref{fig:VZ}.

\begin{figure}[t]
\begin{center} \epsfxsize=130 mm\epsfbox{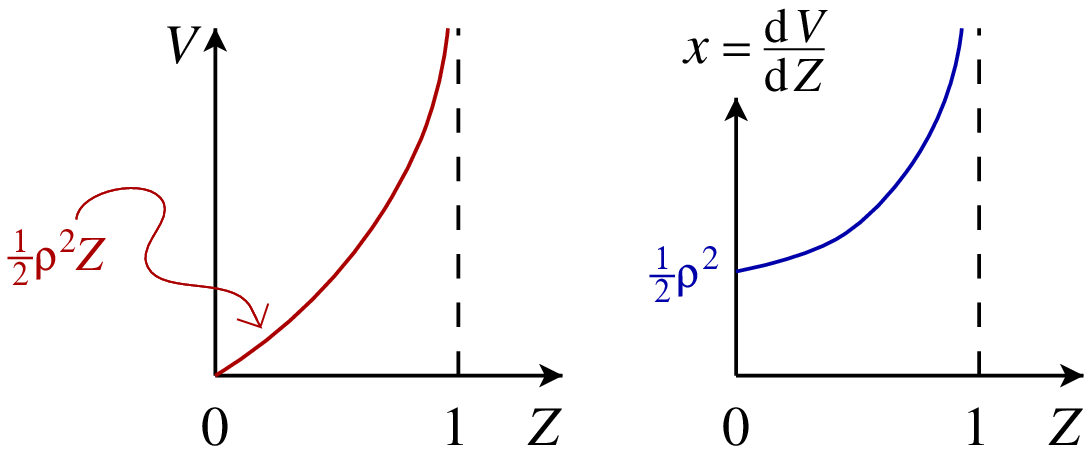}
  \caption{\footnotesize{The relation between \(Z(\phi)\) and \(V(\phi)\)
  producing confinement, and the resulting relation between \(x\) and \(Z\).}}
  \label{fig:VZ}
\end{center}
\end{figure}

With this result we now return to Eq.~\eqn{Lphi}, which we
rewrite as \be\LL=Zx-V\quad,\qquad x=-\quart
F_{\m\n}F_{\m\n}\,.\le{Lx} Extremizing this with respect to
\(\phi\) gives \be\part\LL\phi=0\ \ra\ \part VZ=x=-\quart
F_{\m\n}F_{\m\n}=\half E^2\ .\le{xE2} This allows us to eliminate
\(\phi\) by writing \be\dd\LL=x\dd Z+Z\dd x-\dd V=Z\dd X\ ;\quad
Z={\dd\LL\over\dd x}\ .\le{ZLx} Combining Eq.~\eqn{VZ} with
\eqn{xE2}, we find the relation between \(x\) and \(Z\).
Inverting this, using the relation \(Z(x)\), we find \(\LL(x)\),
and we see that at \be x=\half E^2=\half \r^2\ ,\quad
{\dd\LL\over\dd x}=0\ .\le{extreme} Apparently, the function
\(f(x)\), which in the Maxwell case is just equal to \(x\), must
now have an extremum at the value \eqn{extreme}.

This establishes the function \(f(x)\) for all \(x\ge\half\r^2\).
How exactly the function continues for \(x\) smaller than that
value is of no direct consequence for the confinement mechanism;
of course \(f\) must approach the classical value \(x\) at large
\(|x|\).  Figure~\ref{fig:confcurve} shows the possibilities.

\begin{figure}[t]
\begin{center} \epsfxsize=130 mm\epsfbox{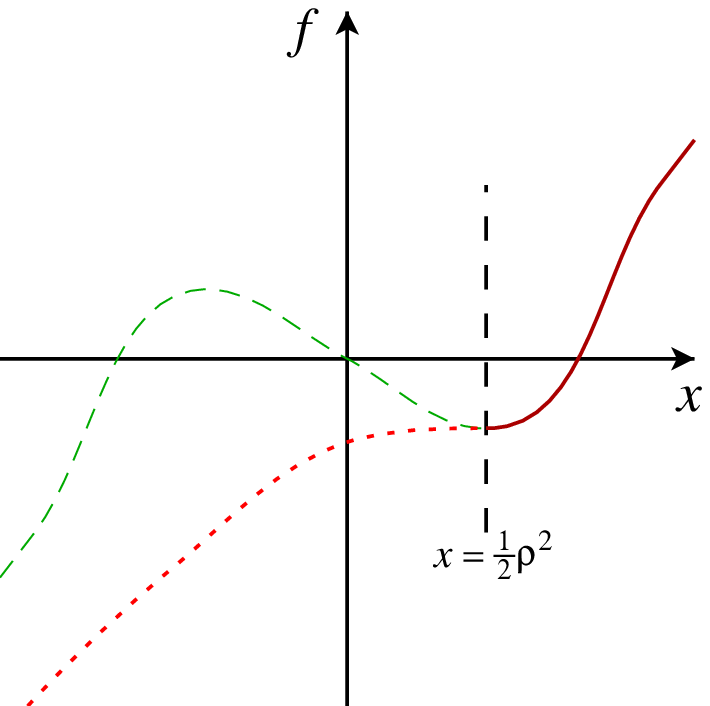}
  \caption{\footnotesize{Different possibilities for the function \(f(x)\)
  yielding confinement. The behaviour to the right of the extremum (solid curve)
  must be as indicated.}}
  \label{fig:confcurve}
\end{center}
\end{figure}

Finally, let us derive a curve \(f(x)\) that would produce
magnetic confinement, \textit{i.e.} the Higgs mechanism. We
perform a dual transformation:\be G_{\m\n}\equiv
Z(\phi)\,\half\,\e_{\m\n\a\b}F_{\a\b}=Z\,\tilde
F_{\m\n}\,.\le{dual} In the source-free case, the Maxwell equation
\(\pa_\m(Z\,F_{\m\n})=0\) implies \be\pa_\a G_{\b\g}+ \pa_\b
G_{\g\a}+\pa_\g G_{\a\b}=0\,,\le{homogeneous} so that one may
write \be G_{\m\n}=\pa_\m B_\n-\pa_\n B_\m\,,\le{Bfield} and the
equations are associated to the Lagrangian \be \LL(B,\f)={-\quart
G_{\m\n}G_{\m\n}\over Z(\f)}-V(\f)\,.\le{Ldual} Electric
confinement for the \(F\) and \(A\) fields implies magnetic
confinement for the \(G,\) and \( B\) fields. We see that \(Z\) is
replaced by \(1/Z\). Plugging in the same equation \eqn{VZ},
\(V=\half\r^2\,Z\), as \(Z\ra 0\), we see a prescribed behaviour
when \(y\equiv \quart G_{\m\n}G_{\m\n}=-x=\half B^2\) is positive:
\be\LL(B)=-{y\over Z}-V\ ;&\label{LdualZV}\\
{\dd\LL\over\dd\f}=0\quad \ra\quad
Z=\sqrt{y\over\half\r^2}\quad;&\ V=\sqrt{\half\r^2\,y}\ ;\\
\LL\ra-\r\sqrt{2y}=-\r |\vec B|\ .&\le{magnstring} The resulting
function \(f(x)=f(-y)\) is sketched in Figure \ref{fig:magnconf}.

\begin{figure}[t]
\begin{center} \epsfxsize=130 mm\epsfbox{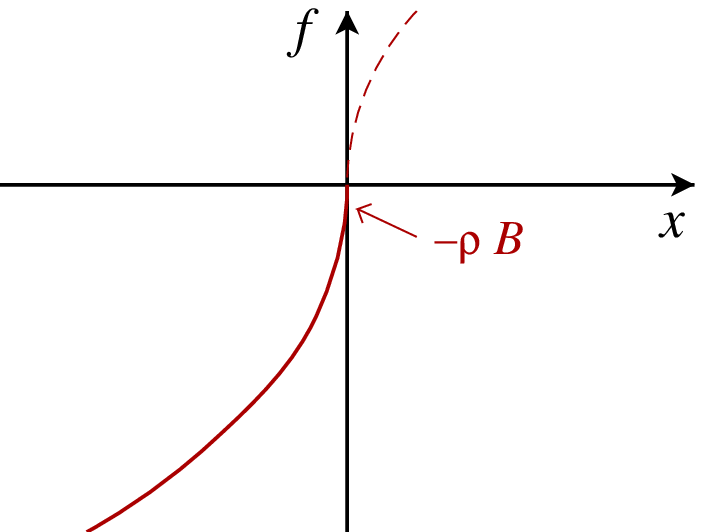}
  \caption{\footnotesize{The function \(f(x)\)
  in the case of magnetic confinement.}}
  \label{fig:magnconf}
\end{center}
\end{figure}

\newsec{Discussion\label{sec:disc}}
Permanent confinement of electric sources is not necessarily a
purely non-perturbative quantum phenomenon. One can choose between
classical, reasonably looking Lagrangians of various kinds that
produce this effect. These Lagrangians are non-renormalizable, and
this means they cannot be used to describe the theory in the far
ultraviolet. We must assume that the soft, infrared divergent
correction terms are due to quantum corrections, but they do not
have to be non-perturbative quantum corrections. The counter terms
used in the first chapters of this contribution, are only slightly
non-local in space, while all are local in time. Arguments that
are very similar to renormalization group arguments can be applied
to conclude that terms of this kind, which become dominating in
the far infrared, necessarily arise when we scale to longer
distances.


\begin{thebibliography}{99}


\bibitem{GrTh} J.~Greensite and Ch.B.~Thorn, {\it Gluon Chain Model of the
Confining Force}, {\tt hep-ph/0112326}.
\bibitem{GtHconf} G.~'t~Hooft, ``Gauge Theories  with  Unified  Weak,  Electromagnetic  and
     Strong Interactions", E.P.S. Int. Conf.  on  High  Energy  Physics,
     Palermo 23-28 June 1975; see also papers reprinted in:
G.~'t~Hooft, ``Under the spell of the gauge principle", Advanced
Series in Mathematical Physics \textbf{19} (1994), Editors: H.
Araki et al. (World Scientific, Singapore), or: G.~'t~Hooft,
\textit{Phys. Scripta} \textbf{24} (1981) 841;
\textit{Nucl.~Phys.} \textbf{B190} (1981) 455.\\ See also: A.M.
Polyakov, \textit{Nucl.~Phys.} \textbf{B120} (1977) 429.

\bibitem{KogutSusskind} Quartic poles were brought in connection
with confining potentials long ago, e.g.: J.~Kogut and
L.~Susskind, \textit{Phys.~Rev.} \textbf{D9} (1974) 3501.

\bibitem{DIAGRAMMAR}  G.~'t Hooft and M.~ Veltman,  "DIAGRAMMAR",
    CERN  Report  73/9  (1973), reprinted in "Particle Interactions at
    Very  High  Energies,  Nato Adv. Study Inst. Series, Sect. B,
    vol.~4b, p.~177.

\bibitem{GtHstrong} G.~'t~Hooft, in Proceedings of the Colloqium
on "Recent Progress in Lagrangian Field Theory and Applications",
 Marseille, June 24-28, 1974, ed. by C.P. Korthals Altes,
  E. de Rafael and R. Stora.

\end{thebibliography}
\end{document}